# Probing infrared excess connection with Li enhancement among red clump giants

Anohita Mallick,[1,2]⋆ Bacham E. Reddy,[1] C. Muthumariappan[1]
[1]*Indian Institute of Astrophysics, 560034, 100ft road Koramangala, Bangalore, India*
[2]*Pondicherry University, R. V. Nagara, Kala Pet, 605014, Puducherry, India*

**ABSTRACT**
We have performed a search among low mass red giants for finding evidence for merger scenario for triggering He-flash and subsequent Li enhancement. We chose a sample of red giants from GALAH survey with well-measured Li abundances, and near- and mid-IR fluxes from 2*MASS* and *WISE* surveys, respectively. The sample contains 418 cool red clump giants and 359 upper red giant branch giants. Most of the giants and majority of super Li-rich giants show no IR excess. Only five red clump giants and one RGB giant show IR excess. Notably, of the five red clump giants with IR excess, three are super Li-rich (A(Li) ≥ 3.2 dex), and two are Li-rich (A(Li) ≥ 1.0 dex). Results suggest Li enhancement among red clump giants may be due to two channels: one resulting from in-situ He-flash in single star evolution and the other due to He-flash triggered by events like merger of He-white dwarfs with giants' He-inert core on RGB. In the latter case, IR excess, as a result of mass loss, is expected from merger events. We have modelled IR excess in all six giants using DUSTY code and derived dust parameters. The estimated kinematic ages and time scales of dust envelopes of the super Li-rich phase suggest Li enhancement took place very recently. Further, the analysis shows a significantly higher proportion (four out of five red clump giants) of rapid rotators ($v_{sini} \geq 8$ km.s$^{-1}$) among Li-rich giants with IR excess compared to Li-normal and Li-rich giants with no IR excess.

**Key words:** stars: evolution — stars: mass-loss — stars: abundances — stars: rotation — surveys

## 1 INTRODUCTION

The canonical stellar evolutionary models predict severe depletion of Li among red giants Iben (1967) and the observations in general agree with the theoretical predictions by Lind et al. (2009). However, observed lithium excess among few red giants has been a long-standing problem since its first discovery in 1980s (Wallerstein & Sneden 1982). Recent studies based on large data sets of stellar spectra from surveys like LAMOST, GALAH, Gaia-ESO and time-resolved photometric survey of Kepler made significant advances in understanding the Li excess origin among low mass red giants. While the spectroscopic surveys enabled to assemble a large number of giants with Li abundance measurements, Kepler photometry and Gaia astrometry made it possible to tag stars accurately to various phases of stars' evolution on the red giant branch (RGB); first dredge-up, luminosity bump, upper RGB, red clump (RC) giants (Deepak & Reddy 2019; Singh et al. 2019; Kumar et al. 2020; Singh et al. 2021; Magrini et al. 2021a; Romano et al. 2021; Magrini et al. 2021b). In particular, asteroseismic analysis based on Kepler photometric data made it possible to unambiguously separate giants ascending RGB with inert He-core surrounded by hydrogen burning shell from those of red clump giants of He-core burning phase (Bedding et al. 2011). This was not possible, earlier, due to ambiguity caused by overlapping positions of giants in RC and RGB phases in the $T_{\text{eff}} - L$ plane of Hertzsprung-Russel (HR) diagram.

Recent studies provided strong evidence that the Li excess phenomenon is ubiquitous among low mass giants of the red clump region and demonstrated that Li abundance only depletes among giants while ascending the RGB (Kumar et al. 2020). As a result, the search for the origin of Li excess significantly narrowed down to the red clump region. Further, the study by Singh et al. (2021) provided credible evidence that Li enhancement origin lies mainly during the short period of the He-flashing phase of about 2 Million years. They showed that most of the super Li-rich (A(Li) ≥ 3.2 dex) stars are exclusively young red clump giants and the Li-poor ones are relatively older red clump giants. Also, studies of Li abundances in open clusters based on Gaia-ESO survey (e.g Magrini et al. 2021b)) indicate Li production among red clump giants. Inspired by these observational results a couple of theoretical models emerged to explain Li excess among RC giants. The models predict high levels of Li abundance either during the main He-flash, at the start of He-ignition at the core (Schwab 2020), or at the tip of RGB (Mori et al. 2021). The initial models are based on certain adhoc assumptions, which needs to be probed further in detail.

It is reasonably established that the He-flash holds the key for both the production and quick transport of Li, produced in hot interiors, to the photosphere. Also, the recent extensive studies based on large samples ruled out the earlier suggestions of direct injection of Li via mergers as most of the Li-rich giants are confined to only RC phase.

⋆ E-mail: anohita.mallick@iiap.res.in

Observations seem to confirm the theoretical predictions in which models require mixing of material of undiluted Li by merging of hundreds of planets (Carlberg et al. 2016) simultaneously, which is quite unlikely. In the case of brown dwarfs, studies limit maximum levels of enhancement of A(Li) ~ 2.6 dex, which is much less than what is observed in some of the giants (Aguilera-Gómez et al. 2016).

Some studies do suggest triggering of Li production in stars by large scale events such as mergers of He-white dwarfs (Zhang & Jeffery 2013; Jeffery & Zhang 2020). The merger events, depending on white dwarf size can trigger He-flash at the core of the star leading to the production of Be and subsequent transportation to cooler regions. Such merger events probably could happen anywhere on RGB, resulting in the RGB giant converting to a red clump giant with He-core burning at the centre. The merger studies also predict ejection of stellar material and high stellar rotation due to angular momentum transfer from the companion. The ejected material can form dust grains as it cools down, and the dust present in the circumstellar regions can show significant infrared excess.

In this study, we have assembled a large sample of upper RGB giants ascending towards the RGB tip and red clump giants, post-He-flash and searched for a correlation between infrared excess and Li abundance among sample giants. Further, mass ejection due to the merging of a sub-stellar object in a giant's atmosphere is expected preferentially to be in the equatorial regions. Hence, we aim to address the geometry of the circumstellar mass distribution using mid-IR flux ratios for those giants that are found to have IR excess for providing possible evidence of merger events with the red giant.

## 2 SAMPLE

For this study, we used a sample of giants of both red clump and upper RGB stars for which well-measured (no upper limits) values of Li abundances, infrared photometry and parallaxes are available. The primary sample is taken from the recently released GALAH (Galactic Archaeology with HERMES) survey DR3 data (Buder et al. 2021). GALAH is a high resolution (R=28000) stellar optical spectroscopic survey with HERMES spectrograph mounted on a 3.6 m Anglo-Australian telescope at Mt. Stromlo Siding Spring Observatory. The GALAH survey DR3 catalogue contains derived stellar parameters and abundances of 30 elements for more than half a million stars. Conveniently, DR3 also lists relevant astrometric data (such as parallaxes, G-magnitude etc.) from the European Gaia space mission (Gaia Collaboration et al. 2018). To better constrain the sample in the $L - T_{eff}$ or Hertzsprung-Russel Diagram (HRD) we limited the GALAH sample with $T_{eff}$ uncertainty $\Delta T_{eff} < 100K$, positive parallaxes and the Gaia fractional parallax error f ≤ 0.15 (= $\frac{\sigma_\omega}{\omega}$, a ratio of 1$\sigma$ parallax uncertainty to the parallax). GALAH catalogue also provides a quality index or flag against each measured quantity.

The quality index of the measured stellar parameters ($T_{eff}$ and log g) can have a value between 0 and 1024; the value 0 refers to the measurement with the least uncertainty and the value 1024 refers to the highest uncertainty in the measurement. The above selection criteria yielded a sample of 302,227 stars.

Since the Li enhancement is known to be prevalent only among the low mass giants we only chose giants with mass ≤ 2 $M_\odot$. Mass of a giant was estimated from its luminosity ($L$), surface gravity ($g$) and effective temperature ($T_{eff}$) using the relation

$$\frac{M}{M_\odot} = 10^{\left[\log\left(\frac{L}{L_\odot}\right) + \log g - \log g_\odot + 4 \times \log\left(\frac{T_{eff\odot}}{T_{eff}}\right)\right]}$$

where Log $g_\odot$ = 4.44 dex and $T_{eff\odot}$ = 5772 K are adapted as the solar values. Luminosity for stars were derived using Gaia parallaxes and G-magnitude using the relation

$$-2.5 \log\left(\frac{L}{L_\odot}\right) = M_G + BC_G(T_{eff}) - M_{bol\odot}$$

where the absolute magnitude in G-band is given as

$$M_G = G + 5 - 5 \log_{10} r - A_G$$

and the temperature dependent bolometric correction is

$$BC_G(T_{eff}) = \sum_{i=0}^{4} a_i (T_{eff} - T_{eff\odot})^i$$

Here $a_i$s are the polynomial coefficients of the model function as defined above. Values of $a_i$s are taken from Andrae et al. (2018). The value of $M_{bol\odot}$ = 4.74 for the Sun, $r$ is the distance to the star (in parsec) and $A_G$ is the interstellar extinction in G-band. Since systematic measurements of $A_G$ are not available for the entire sample of giants, we have not applied extinction corrections in this study. More details on the extinction correction are given in section 3.3. Using the cut-off mass of ≤ 2$M_\odot$, we made a sample of 242, 431 giants. The entire sample is shown in the HR diagram (Figure 1).

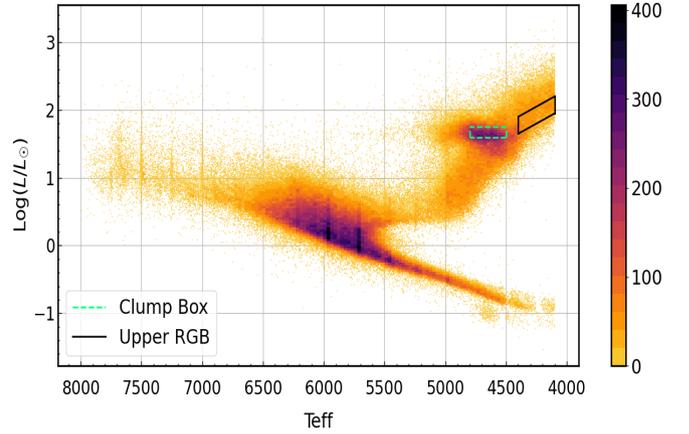

**Figure 1.** Selection of sample giants using GALAH data in HR-diagram (red clump: dotted green box, upper RGB: black box)

To choose sample of giants just before and after the RGB tip we drew two boxes, one for the RC sample and another for upper RGB, with the following range in $T_{eff}$ and $L$; 4500 K ≤ $T_{eff}$ ≤ 4800 K and 1.6$L_\odot$ to 1.75$L_\odot$ for RC, and 4100 K ≤ $T_{eff}$ ≤ 4400 K and 1.65 ≤ $L_\odot$ ≤ 2.2$L_\odot$ for the RGB sample. These definitions resulted in a sample of 8808 RC and 3607 upper RGB giants. Further, to ensure having giants with well measured Li abundances we adopted only giants with `flag_Li_fe`= 0 and `flag_fe_h` =0 as Li in the GALAH is given with respect to star's Fe abundance. Upper limits (`flag_Li_fe` = 1) were not considered. Finally, we have a sample of 957 RC and 518 RGB giants for which well measured stellar parameters and Li abundances are available.

We used IR data from infrared photometric surveys of 2*MASS* and *WISE* missions to understand if there is a relation between Li abundance and IR excess. While 2*MASS* provides photometric survey of stars in near-infrared bands J(1.235 $\mu$m), H (1.662 $\mu$m) and K (2.159 $\mu$m), *WISE* provides photometric data in their mid-IR bands W1 (3.3526 $\mu$m), W2 (4.6028$\mu$m), W3 (11.5608 $\mu$m) and W4 (22.0883 $\mu$m). All the giants in our final sample of RC and RGB have 2*MASS* fluxes (referred as sample 1), but only 418 RC and 359 RGB giants have both *WISE* and *2MASS* fluxes ( referred as sample 2). We have not included the giants with flux values provided as upper limits.

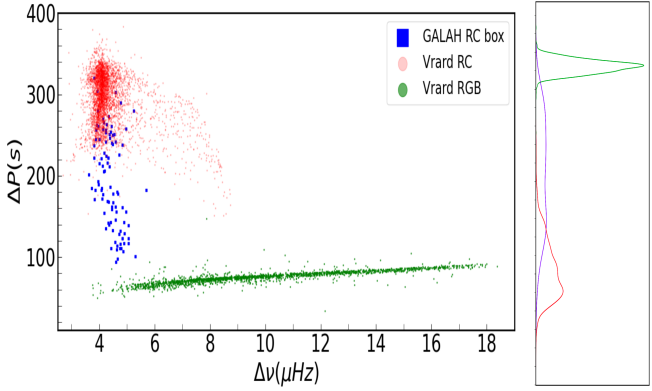

**Figure 2.** Estimation of overlap between the RC and RGB sample in our GALAH sample selection using RC (red) and RGB (green) samples by Vrard et al. (2016) for which evolutionary phases are determined based on asteroseismic analysis. The 96 giants (blue ) in our GALAH sample for which classification is available based on secondary calibration are shown.

## 3 ANALYSIS AND RESULTS

### 3.1 RGB contamination in Hertzsprung-Russell diagram

In Fig. 1, the colour bar indicates the number density of stars; darker the colour higher the number density of stars. The sample was chosen so that the selected RC giants will have minimum contamination with the RGB sample. However, given the uncertainties in stellar parameters and a small difference in $T_{\rm eff}$ and log $g$ between the giants of RC and RGB, a small overlap between the two samples cannot be ruled out. To estimate the level of contamination, we compared our sample of RC giants with a sample of giants from Vrard et al. (2016) for which evolutionary phase was determined using asteroseismic analysis. The Vrard et al. sample is plotted in a seismic plot of g-mode period spacing ($\Delta P$) and average large-frequency spacing ($\Delta \nu$) (see Fig. 2). RGB giants (green symbols) are well separated from RC giants ( red symbols) in $\Delta P$ but have overlap in frequency at the lower end. Unfortunately, none of our selected sample giants from GALAH are common with the Vrard et al. (2016) sample. However, we found 96 of our RC sample giants common with a much larger sample of half a million RC giants in the catalogue (Lucey et al. 2020) for which $\Delta p$ and $\Delta \nu$ values are inferred based on secondary calibration using a neural network from stellar SEDs and spectra from LAMOST DR3 and APOGEE DR14 (Ting et al. 2018). The 96 common RC giants (blue symbols) from our sample are compared with the Vrard et al. sample in Fig. 2. As shown in Fig 2, RGB giants (green) show an average $\Delta P$ of $\sim$ 80 seconds with a dispersion of $\sim$ 20 seconds, suggesting giants with $\Delta P >$ 100s could more likely be RC giants. We found only 4-5 giants, out of 96 RC giants (blue), from our GALAH sample overlapping with the RGB giants (green). This provides a rough indication of about 5% RC contamination with RGBs in the present sample. A similar exercise was done by Kumar et al. (2020) who estimated the contamination of GALAH RC with RGB at about 10%. The difference in the contamination between the two samples may be attributed to tighter box definition for the RC

| Group | RC (%) | RGB (%) | Total |
|---|---|---|---|
| Sample 1 (having *2MASS* fluxes) | | | |
| Li-normal | 629 (65.7) | 469 (90.5) | 1098 |
| Li-rich | 306 (31.9) | 47 (9) | 353 |
| SLR | 22 (2.3) | 2 (0.4) | 24 |
| Sample 2 (having both 2MASS and *WISE* fluxes) | | | |
| Li-normal | 241 (57.6) | 326 (90.8) | 567 |
| Li-rich | 164 (39.2) | 32 (8.9) | 196 |
| SLR | 13 (3.11) | 1 (0.29) | 14 |

**Table 1.** Distribution of RC and RGB stars from sample 1 and 2

sample selection and a more extensive and improved calibration of the GALAH DR3 sample.

### 3.2 Lithium Abundance

GALAH DR3 catalogue provides Li abundances derived from spectral synthesis of Li resonance line at 6707.8Å. The abundance in the catalogue is given in the form of $[Li/Fe]$ [1]. Generally, the Li abundance in the literature is expressed as

$$A(Li) = 12 + \log_{10} \frac{n(Li)}{n(H)}$$

where, n(Li) is no. of Li atoms and n(H)=$10^{12}$ is the assumed number density of H atoms. By using the GALAH adopted solar Li abundance of A(Li)$_\odot$ = 1.05 $\pm$ 0.10 dex and solar metallicity [Fe/H]$_\odot$ = 7.50 $\pm$ 0.04 (see Buder et al. (2021)) Li abundances for the sample stars are extracted using the expression given below :

$$A(Li) = [Li/Fe] + [Fe/H] + A(Li)_\odot$$

Uncertainty in A(Li) is given as:

$$\sigma_{A(Li)} = (\sigma^2_{[Li/fe]} + \sigma^2_{[fe/h]} + \sigma^2_{A(Li)_\odot})^{1/2}$$

$\sigma_{[Li/fe]}$ and $\sigma_{[fe/h]}$ are propagated from GALAH and $\sigma_{A(Li)_\odot}$ = 0.10 dex ( uncertainty on $A(Li)_\odot$ )

For our sample, $\pm$ 0.11 $\leq \sigma_{A(Li)} \leq \pm$ 0.23. As discussed in section 2, stars with only reliable Li and Fe measurements have been selected. The distribution of RC and RGB samples are shown in Fig 3. RC sample peaks at about A(Li) = 0.62 dex, and for RGB, peak is at $\sim$ 0.15 dex. The Li abundance distributions are, in general, similar to that shown by Kumar et al. (2020); Mori et al. (2021) although some of the lower A(Li) stars in this study might have been dropped out while selecting good infrared photometric data.

For convenience, following the recent study by Singh et al. (2021), we have divided the sample into three groups based on Li abundances. The groups are

(i) Li-normal giants, A(Li) < 1.0 dex
(ii) Li-rich giants, A(Li) $\geq$ 1.0 dex
(iii) Super Li-rich giants or SLRs with A(Li) $\geq$ 3.2 dex

Number statistics for each group along with their percentages are provided in Table 1.

Note that Li abundances of few RGB giants with A(Li) $\geq$ 1.8 dex,

---

[1] $[Li/Fe]$ defined as $[Li/Fe] = [Li/H] - [Fe/H]$ where $[Li/H]$ = A(Li)$_\star$ - A(Li)$_\odot$ and [Fe/H] = A(Fe)$_\star$ - A(Fe)$_\odot$.

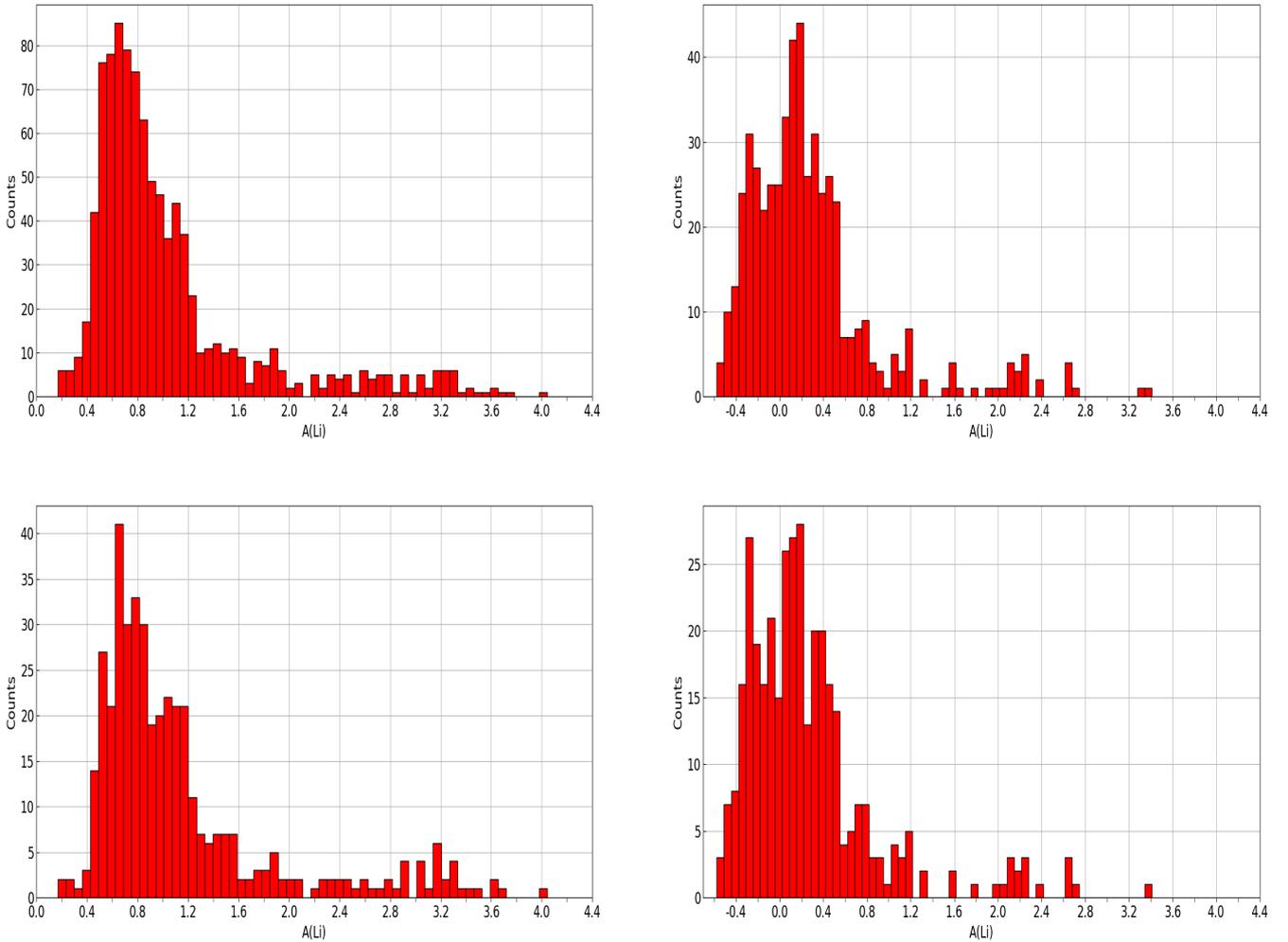

**Figure 3.** Distribution of Li abundances among our sample RC and RGBs. In top panels we showed Li distribution for the entire sample of RC (left) and RGB (right) and in the bottom panels for RC giants (left) and RGB (right) with mid-IR colours.

as shown in the distribution, are at odds with theoretical models and most of the observational studies (Lind et al. 2009; Kumar et al. 2020) which demonstrated that Li abundance gets destroyed among RGB stars, reaching as low as A(Li) ∼ −1.0 dex before reaching the tip. Probably, these Li-rich RGB giants are a result of contamination with RC giants. Also, this could be due to uncorrected and differential extinction from star to star (see Deepak & Reddy (2019)). Since we do not have extinction values for the entire sample we have not applied extinction corrections, and will not be dealing, in this study, with the exact nature of these few Li-rich RGB giants' evolutionary phases. Probably, these may be RC giants. It would be worthwhile to conduct asteroseismic analysis to know the exact evolutionary status. At present, many of these giants do not have asteroseismic data.

### 3.3 Near and Mid-Infrared Excess

It was hypothesised that the Li enhancement in red giants is linked to mergers with He-white dwarfs (Zhang & Jeffery 2013; Jeffery & Zhang 2020). It is also predicted that such mergers result in the ejection of material, and subsequently causing excess emission in the infrared. To investigate the IR properties of the sample stars, as described in section 2, we have used the IR photometric measurements taken from Two Micron All Sky Survey (2*MASS*; Cutri et al. (2003)) and Wide-field Infrared Survey Explorer (*WISE*; Cutri et al. (2021)). Presence of IR excess is the evidence for the presence of circumstellar dust around a star whose properties can be traced using a two-colour diagram, and fitting spectral energy distribution (SED). The colour-colour diagrams of RC and RGB sample in *WISE* and 2*MASS* bands are shown in Fig. 4.

The spread in the IR colour-colour diagram is due to the observed range in $T_{\text{eff}}$ and metallicity of the sample giants (Wright et al. 2008; Alonso et al. 1994). The large spread in *WISE* colour along the X-axis compared to Y-axis may be due to a steep decrease in fluxes at longer wavelengths than at shorter wavelengths. We have superposed the sample data with zero IR excess boxes to identify IR excess stars among the sample. The theoretical boxes are constructed by calculating the colours using the sample upper and lower limits of $T_{\text{eff}}$. The theoretical colour between two wavelengths is given by

$$[\lambda_1 - \lambda_2] = -\log\left[\left(\frac{\lambda_1}{\lambda_2}\right)^\alpha \left(\frac{B_{\lambda_1}}{B_{\lambda_2}}\right)\left(\frac{f_{\lambda_2}}{f_{\lambda_1}}\right)\right]$$

where $B_{\lambda_1}$ and $B_{\lambda_2}$ are Planck functions at wavelengths $λ_1$ and $λ_2$ for a given $T_{\text{eff}}$, respectively. Terms $f_{\lambda_1}$ and $f_{\lambda_2}$ are the zero point flux densities at the corresponding wavelengths. The spectral index ($\alpha$) is the deviation from the blackbody spectrum and is varied manually to cover all the zero excess stars. Value of $\alpha$ varied from -1 to 3 for *WISE* plots and -1 to 1 for *2MASS* plots. As revealed by the plots in Fig 4, there are just six giants outside the zero excess boxes; five RC giants and one RGB giant, indicating they have IR excess. None of the giants in sample 1 show IR excess. All six giants are in sample 2 showing evidence for IR excess in *WISE* colours. Further analysis of IR excess is given below in section 3.4.

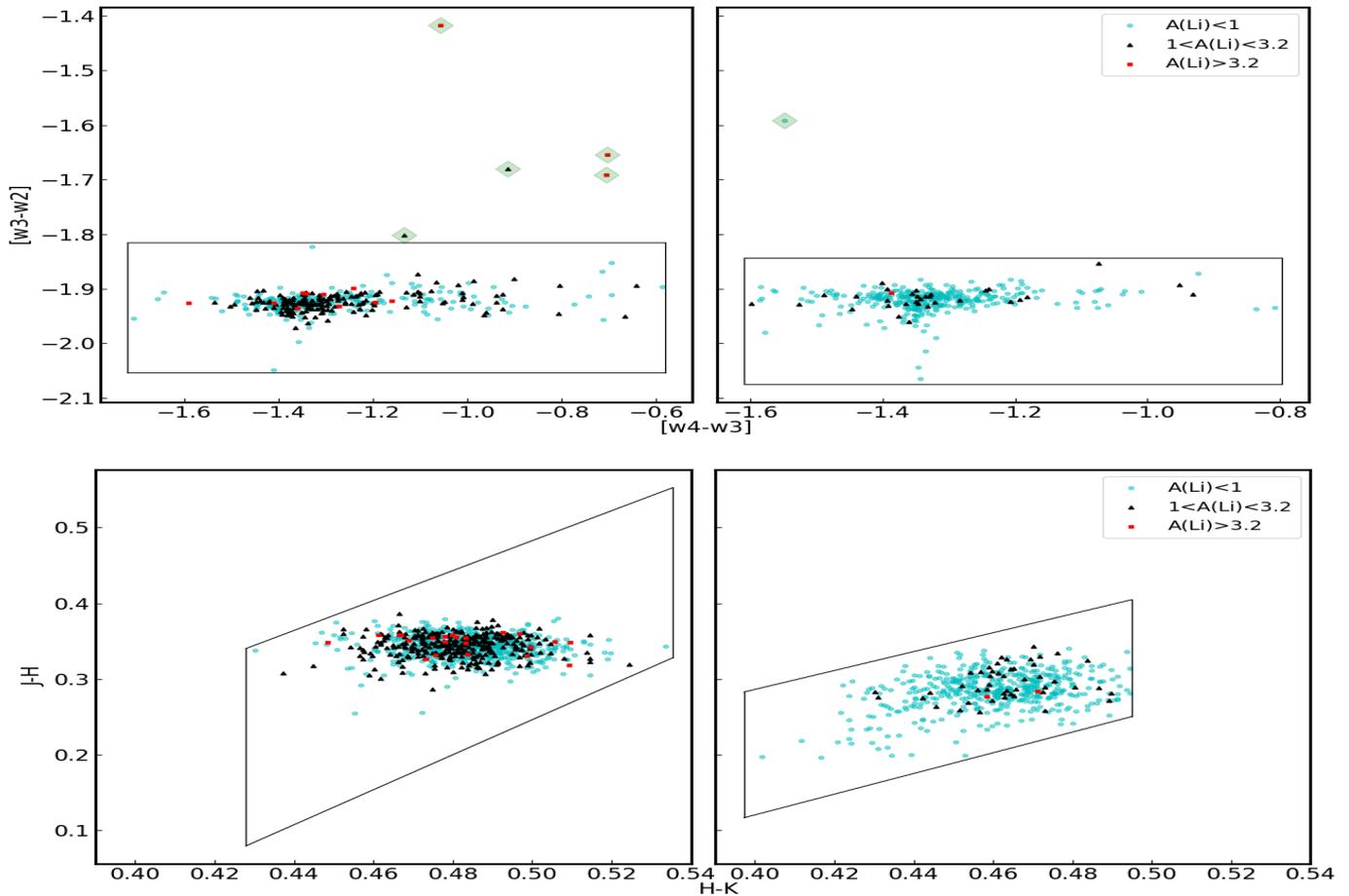

**Figure 4.** Colour-Colour Diagrams for *WISE* mid-IR (top panels) and for *2MASS* near-IR colours (bottom panels). Left and right panels show the RC and RGB giants, respectively. The boxes cover no IR-excess regions in CCDs. Note, only six giants in mid-IR colour panels are outside the boxes i.e they show IR excess.

As said earlier, we have not corrected the sample for interstellar extinction as extinction values ($A_G$) are not available for many of the giants. Large extinctions may lead to erroneous designation of stars' evolutionary phase. Luckily, all the six giants that have IR excess have $A_G$ estimations from the Gaia catalogue. Maximum correction is for the upper RGB star BD-19 4687 with $A_G$= 0.98. This results in an uncertainty of 0.08 in $\log\left(\frac{L}{L_\odot}\right)$ and 0.08 in mass. As shown in Fig. 5, the evolutionary state of none of the six giants changes after applying extinction correction. .

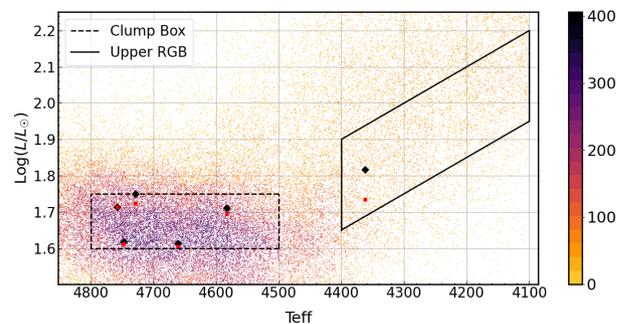

**Figure 5.** Sample stars in the H-R diagram showing upper RGB and RC regions. Six giants for which IR excess have been shown before (red squares) and after the extinction correction (black diamonds). Note, evolutionary status of all the six giants remains unchanged.

### 3.4 Spectral Energy Distribution (SED) of IR excess stars

For the six giants that show evidence for IR excess in *WISE* colour-colour diagrams (CCDs), we have constructed SEDs from the observed photometry in optical and IR bands and compared them with the model SEDs. The observed SED includes contributions from both the stellar photosphere and the dust shell around the star. We have used Castelli and Kurucz 2004 Stellar Atmosphere Models, which has a collection of 4300 LTE photospheric stellar models with varying $T_{eff}$, log$g$ and [Fe/H]. Fig 6 shows the SEDs of all six IR excess stars along with their photospheric contribution from model atmosphere. $T_{eff}$, log$g$ and [Fe/H] needed for stellar model atmosphere are taken from GALAH DR3. Optical, Near-IR and Mid-IR flux densities are taken for each star using the filter repository of VOSA (Virtual Observatory SED Analyzer), developed by Spanish Virtual Observatory project. Relevance of the WISE band excess for these six giants is discussed further in section 4.

We used 1-D radiative transfer code DUSTY (Ivezic et al. 1999) to fit the observed SEDs of sources which show IR excess. We assumed that the dust shell is spherically symmetric, originating from a point source at the center. The following input parameters are supplied to obtain the best model fit to the observed SED.

- Input flux is the model stellar spectrum obtained from Kurucz models for a given $T_{eff}$, log$g$, [Fe/H].
- Dust properties: we considered astronomical silicates for the circumstellar dust and their optical constants are inbuilt in DUSTY.
- Grain size distribution - standard Mathis, Rumpl, & Nordsieck (1967) - MRN distribution with $a_{min}$ = 0.005 $\mu$m and $a_{max}$ = 0.25 $\mu$m
- dust temperature ($T_{inner}$) at the inner radius $R_{in}$ of the envelope
- The relative thickness $\left(\frac{R_{out}}{R_{in}}\right)$ of dust shell
- optical depth ($\tau_v$) for a fiducial wavelength $\lambda_0 = 0.55 \mu$m.

Envelope radial density distribution was computed by DUSTY for the case of radiatively driven winds by varying input flux, $T_{inner}$, $\left(\frac{R_{out}}{R_{in}}\right)$ and $\tau_v$. The modelled SED is matched with the observed SED by scaling the model flux to the observation at 2$MASS$ K-band.(see Fig 6). As shown in the figure, the model SEDs fit quite well with the observed SEDs in the wavelength range from optical to far-IR. The best fit model parameters for the IR excess giants are given in Table 3.

### 3.5 Mass loss and kinematics of dust envelope

Mass loss rates and kinematic age are derived using the formula (de La Reza et al. 1996) by combining the derived parameters from the DUSTY and the values taken from Chan & Kwok (1988). The mass-loss rate is calculated as

$$\dot{M} = \frac{16\pi}{3}\left(\frac{\rho_d}{\psi}\right)\left(\frac{a}{Q_\lambda}\right)\tau_v V_s R_{in}$$

Kinematic age of dust shell is given by

$$t_d = \frac{R_{in}}{V_s}$$

where, $\rho_d$ is the grain density (3 $g.cm^{-3}$), $\psi$ is the dust to gas mass ratio (4.3×10$^{-3}$), $Q_\lambda$ is the continuum dust absorption coefficient, $a$ is the radius of a grain, $\tau_v$ is the optical depth at visible band and $V_s$ is the wind velocity (2 km/s). The value of $\left(\frac{a}{Q_\lambda}\right) = \frac{\lambda}{12.6}$ $\mu$m in the thermal IR region. The derived value of $R_{in}$ from the DUSTY code is for a stellar luminosity of $10^4$ L$_\odot$. By scaling ($R_{in} \propto L^{1/2}$) the derived $R_{in}$ for the actual luminosity of IR excess giants and substituting the values of the parameters, the mass-loss relation is simplified as

$$\dot{M} = 1.62 \times 10^{-22} \tau_v R_{in} \ M_\odot.yr^{-1}$$

The derived mass-loss ($\dot{M}$) by fitting the DUSTY and kinematic age ($t_d$) are given in Table 3.

The values of $\dot{M}$ derived from the DUSTY for all the giants are compared with the respective mass-loss rates derived for a given set of stellar parameters using the modified Reimers' Law (Schröder & Cuntz 2005)

$$\dot{M}_R = \frac{\eta L_* R_*}{M_*}\left(\frac{T_{eff}}{4000}\right)\left(1 + \frac{g_\odot}{4300 g_*}\right)$$

Stellar parameters ($L_*$, $R_*$, $g_*$ and $T_{eff}$) are taken from the Table 2 and $M_*$ was calculated using the relation in Sec. 2. We have adopted a value for the fitting parameter $\eta = 8 \pm 1 \times 10^{-14} M_\odot.yr^{-1}$ (Schröder & Cuntz 2005) and log $g_\odot$ = 4.44 dex. The uncertainties in mass loss are estimated using uncertainties in the Gaia parallaxes. As shown in Table 3, mass-loss rates derived from DUSTY modelling are 2-4 magnitudes higher than the expected mass-loss rates from the Reimer's relation for the giants' evolutionary phase. This suggests that the giants with IR excess might have, in the recent past, experienced some kind of merger events resulting severe mass loss. The derived values of $t_d$ of dust shells give the ages for the circumstellar envelopes < 4000 years, see Table 3.

### 3.6 Asymmetrical distribution of circumstellar dust

Results obtained from DUSTY are based on the assumption that the dust shell has spherical geometry with symmetric mass loss. The assumption may be valid if the mass loss is due to evolutionary effect and caused by stars' internal mechanism. However, if the mass loss is due to external events like mergers of white dwarfs one would expect asymmetric mass loss resulting in dust shells of non-spherical geometry; torus or disk shaped geometry for the circumstellar mass distribution (Lynden-Bell & Pringle 1974). To understand the geometry of dust shells around these stars we used the following relation to fit the observed WISE colours:

$$\frac{F_{\lambda_1}}{F_{\lambda_2}} = \left(\frac{\lambda_2}{\lambda_1}\right)^\beta \frac{B_{\lambda_1}(T_d)}{B_{\lambda_2}(T_d)}$$

where $B_{\lambda_1}(T_d)$, $B_{\lambda_2}(T_d)$ are Planck functions at wavelengths $\lambda_1$ and $\lambda_2$, $T_d$ is the mean dust temperature, and $\beta$ is the opacity index. The value of $\beta$ is an index related to the IR continuum emission flux ratios and is expected to be in the range of 1 to 2 for the spherically symmetric dust shells with sub-micron sized grains (Muthumariappan et al. 2006). The index $\beta$ is different from that of the spectral index, $\alpha$ used in calculating SED (see section 3.3) which are related as;

$$\alpha = \beta + 2$$

By fitting the function with the observed continuum $WISE$ flux ratio of $W2/W3$ and $W3/W4$, we get the mean values for $\beta$. We derived values of $\beta$ for all the IR excess giants in this study and also IR excess giants studied by Bharat Kumar et al. (2015). Values of $\beta$ are given in Table 4. At the IR wavelengths extinction may be very little or nil. The derived $\beta$ values, in general, are close to the value of opacity index for ISM - $\beta_{ISM}$ = 1.8 ±0.2 (Draine 2006) suggesting giants' dust shells are symmetric as viewed in general with sub-micron size grains. Grains are silicates in the dust shells whereas in ISM it is a mixture of silicates with carbonaceous grains. However, two giants in our study do show extremely low values of $\beta$ = 0.04 for BD-19 4687 and $\beta$ = −1.42 for BD+01 143. The low values are attributed to asymmetric or torus like dust distribution as discussed by Lynden-Bell & Pringle (1974).

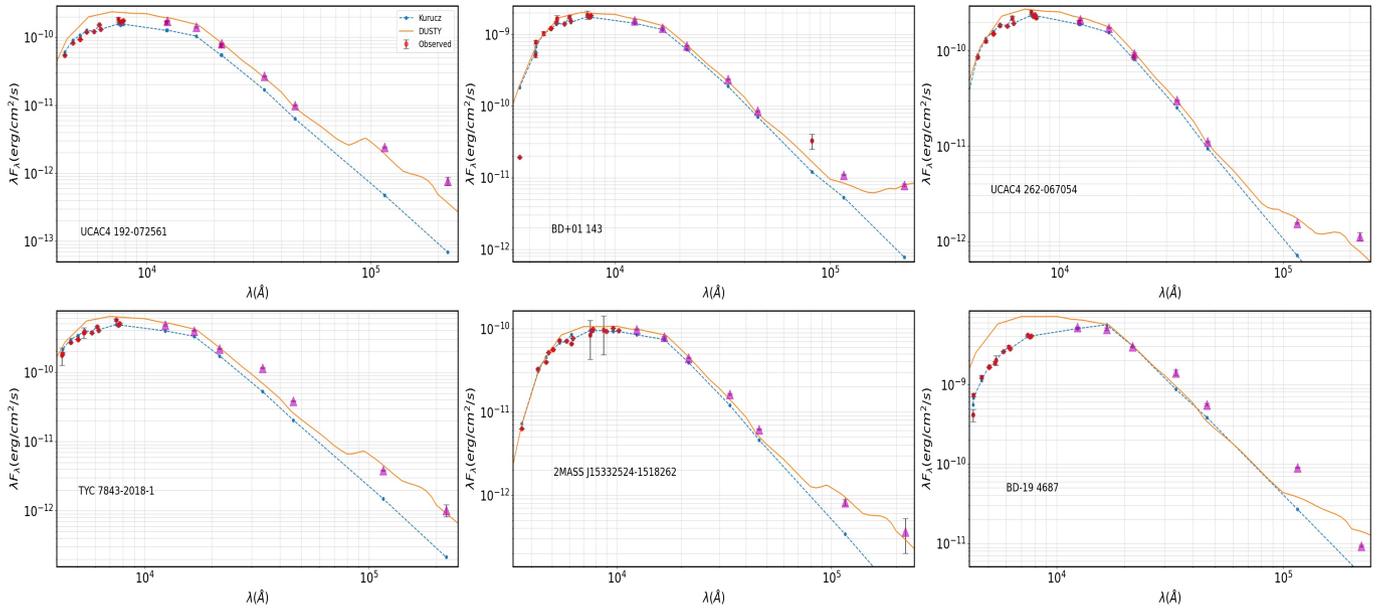

**Figure 6.** Spectral energy distributions of giants with IR excess fitted with DUSTY models. Stellar atmospheric SEDs are from Kurucz models. Triangles indicate IR bands (2$MASS$: J, H, K, and $WISE$: W1, W2, W3, W4)

| Star | $T_{eff}$ (K) | logg | Luminosity ($L_\odot$) (extinction corrected) | Radius ($R_\odot$) (extinction corrected) | A(Li) (dex) | Classification |
|---|---|---|---|---|---|---|
| UCAC4 192-072561 | 4661.60 | 2.34 | 41.01 | 10.06 | 4.04 | RC |
| BD+01 143 | 4747.72 | 2.41 | 40.93 | 9.29 | 3.66 | RC |
| UCAC4 262-067054 | 4757.81 | 2.45 | 51.78 | 10.63 | 3.31 | RC |
| TYC 7843-2018-1 | 4728.81 | 2.31 | 56.14 | 12.05 | 1.59 | RC |
| 2MASS J15332524-1518262 | 4583.01 | 2.29 | 49.55 | 10.67 | 1.41 | RC |
| BD-19 4687 | 4362.29 | 1.80 | 65.56 | 16.68 | 0.29 | RGB |

**Table 2.** Stellar parameters of sample giants with IR-excess detected

| Star | $T_{inner}$ (K) | $\tau_v$ | $T_d$ (K) | $R_{in}$ (cm) | $t_d$ (years) | $\dot{M}$ ($M_\odot.yr^{-1}$) (DUSTY) | $\dot{M}_R$ ($M_\odot.yr^{-1}$) (Reimer's Law) |
|---|---|---|---|---|---|---|---|
| UCAC4 192-072561 | 500 | 4.00E-2 | 194 | 5.92E14 | 93.85 | 3.83 ± 0.03E-9 | 5.19 ± 0.39E-11 |
| BD+01 143 | 100 | 4.12E-2 | 52 | 2.50E16 | 3963.22 | 1.67 ± 0.01E-7 | 4.33 ± 0.06E-11 |
| UCAC4 262-067054 | 225 | 3.00E-2 | 107 | 2.91E15 | 461.32 | 1.41 ± 0.01E-8 | 4.70 ± 0.12E-11 |
| TYC 7843-2018-1 | 440 | 2.85E-2 | 179 | 7.40E14 | 117.31 | 3.41 ± 0.09E-9 | 7.58 ± 1.34E-11 |
| 2MASS J15332524-1518262 | 300 | 4.00E-2 | 136 | 1.45E15 | 229.87 | 9.39 ± 0.02E-9 | 5.40 ± 0.39E-11 |
| BD-19 4687 | 150 | 1.25E-2 | 75 | 6.92E15 | 1097.02 | 1.40 ± 0.12E-8 | 2.79 ± 1.17E-10 |

**Table 3.** The derived dust shell parameters from the DUSTY code for the IR-excess giants.

### 3.7 Stellar rotation

There are suggestions in the literature that the excess Li in some of the low mass giants may be due to rotation induced extra mixing (Charbonnel & Lagarde 2010). Also, studies of merger scenario predict high rotation and presence of IR excess (Kamath et al. 2016; Ceillier et al. 2017). To understand whether there is a pronounced relation between the properties of excess Li, presence of dust and high rotation among giants we analysed $v_{sini}$ values of our sample giants. The GALAH catalogue provides velocity values of $v_{broad}$ which is the overall broadening of stellar profiles, the combination of macro-turbulent ($v_{mac}$) and micro-turbulent ($v_{micro}$) velocities. Since the observed line profile broadening is the convolution of different effects of broadening we estimated $v_{sini}$ values of the sample using the relation of quadratic sum;

$$v_{broad} = \sqrt{v_{sini}^2 + v_{mic}^2 + v_{mac}^2}$$

$v_{mic}$ is provided by GALAH and we have assumed an average $v_{mac}$ = 3 km.s$^{-1}$ (Carney et al. 2008). In Fig. 7 the distributions of $v_{sini}$ with A(Li) are shown for RC as well as for RGB stars. In both classes,

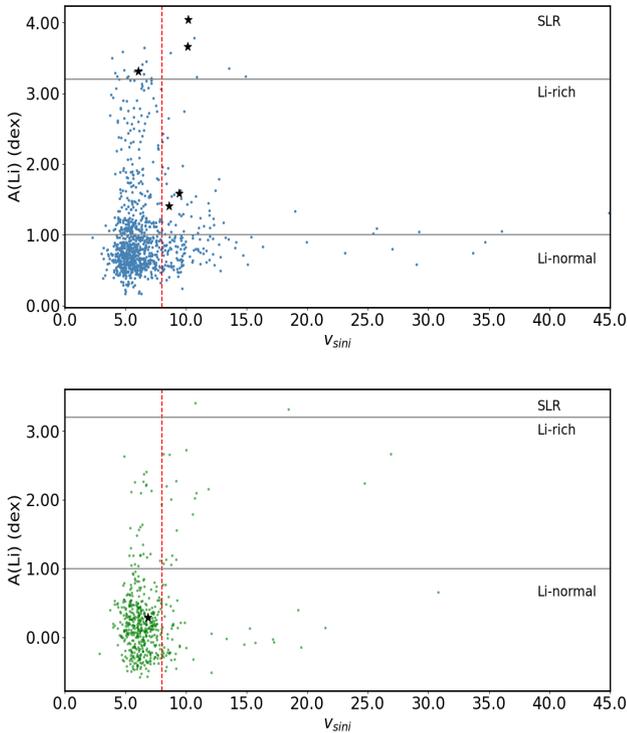

| Star | $\beta$ | $v_{sini}$ km.s$^{-1}$ |
|---|---|---|
| *This work* | | |
| UCAC4 192-072561 | 2.71 | 10.16 |
| BD+01 143 | -1.42 | 10.15 |
| UCAC4 262-067054 | 1.53 | 6.07 |
| TYC 7843-2018-1 | 2.44 | 9.42 |
| 2MASS J15332524-1518262 | 2.12 | 8.61 |
| BD-19 4687 | 0.04 | 6.87 |
| *Kumar et.al. 2015* | | |
| HD 233517 | 1.41 | 17.6 |
| HD 219025 | 2.40 | 23.0 |
| HD 19745 | 2.18 | 3.0 |
| IRAS 13539-4153 | 2.80 | 5.0 |
| IRAS 17596-3952 | 2.48 | 35.0 |
| IRAS 13313-5838 | 2.61 | 20.0 |
| PDS 100 | 2.75 | 8.3 |

**Table 4.** Derived values of opacity index $\beta$ and $v_{sini}$ of IR-excess giants in this study. For the giants in Kumar et al. (2015) only $\beta$ values are derived by us. The $v_{sini}$ values are taken from original studies; HD 233517 (Charbonnel & Balachandran 2000),HD 219025 (Jasniewicz et al. 1999),HD 19745, IRAS 13313-5838,IRAS 13539-4153, IRAS 17596-3952 (Reddy & Lambert 2005), IRAS 13313-5838 (Drake et al. 2002), PDS 100 (Takeda & Tajitsu 2017)

**Figure 7.** Distribution of A(Li) and $v_{sini}$ values among RC (top panel) and RGB (bottom panel) samples. IR excess stars are shown as (★) symbols

most stars lie in the range of 5 < $v_{sini}$ < 10 km.s$^{-1}$. Following Drake et al. (2002), we have considered all stars with $v_{sini}$ > 8 km.s$^{-1}$ to be rapid rotators. For RC stars, the probability of finding a rapid rotator is $p_{(rr,RC)}$ = 0.18 (168 stars in 957) and for RGB stars it is $p_{(rr,RGB)}$ = 0.14 (73 stars in 518) which is similar to the study conducted by (Martell et al. 2021a). However, among giants in sample 2 (with WISE IR colours) probability of finding of rapid rotators among RCs is 0.13 (53/417) and for RGB it is 0.18 (64/357) (see Table 4). There seems to be no distinguishing trend for rapid rotators among RGB and RC giants or between giants with or without mid-IR colours. One would have expected on average higher $v_{sini}$ values for RCs compared to giants on RGBs as giants shrink significantly in size post He-flash. However, four out of 5 RC giants with IR excess are rapid rotators which is about 5-6 times more probable as compared to giants without IR excess. Note, $v_{sini}$ values may be more uncertain given the relatively low resolution and the assumptions we made in the derivation of $v_{sini}$ values. It is important to measure $v_{sini}$ values at least for the IR-excess giants as well as Li-rich stars using very high resolution spectra ( $R \geq 60{,}000$).

## 4 DISCUSSION

A few earlier studies attempted to find whether a correlation exists between the values of Li, IR excess and $v_{sini}$ among giants and an evidence for the external origin of Li-excess in giants. For example, the study by Rebull et al. (2015) analysed a sample of Li-rich giants collected from the literature, and Bharat Kumar et al. (2015) performed a study using a large and unbiased sample. Both the studies concluded IR excess among giants is very rare and found no correlation between Li and IR excess or $v_{sini}$ values. The recent study by Martell et al. (2021a) using the GALAH sample is similar to ours but differ in the analysis. Unlike Martell et al. (2021a), our sample forms a unique set of giants, covering evolutionary phases just before and after the He-flash event and having well-measured near-IR and mid-IR fluxes from 2*MASS* and *WISE* surveys along with Li abundances from the GALAH survey.

The mass loss, or IR excess, could be either due to He-flashes in single stars' evolution or due to mergers that trigger He-flash (Castellani & Castellani 1993; Zhang & Jeffery 2013). In both the cases, excess in near-mid IR colours will be detected if the material was ejected recently. We do not have reliable data in the far-IR colours to probe cool dust. None of the giants, RC or RGB, with 2*MASS* colours (sample 1) shows an excess in near-IR colours (Fig 4, bottom panel), implying no hot dust component around the stars. In the case of sample 2 giants with Mid-IR colours, we found a total of six giants with IR excess, five RC and one RGB, i.e., 0.8% giants with IR excess. Our results are similar to the findings of 0.5% giants by Bharat Kumar et al. (2015) and 0.8% by Martell et al. (2021a) who inferred IR excess using a single colour (W1-W4).

Broadly, results suggest IR excess among giants is very rare. However, if we look at IR excess separately among RC and RGB giants, giants in RC seems to have a much higher probability of having IR excess than those on the RGB phase. In our study, there are five out of 418 RC giants with IR-excess, which is 1.2% against 0.3% of IR excess giants among RGB. Martell et al. (2021a) survey reported 1.4% giants with IR excess among RC, which is similar to ours, but their 1.3% RGB giants with IR excess is significantly larger than ours. Probably, the difference between ours and Martell et al. (2021a) may be to do with the difference in the RGB sample selection. Unlike only upper RGB giants, in our case, their RGB sample includes all the giants starting from the base of the RGB to the RGB tip and also, their sample includes more massive giants. It is not clear whether the IR excess among more RGB giants in their sample is due to the presence of more massive giants which evolve relatively faster and retained some dust from their pre-main-sequence evolution. Interestingly, all the five RC giants with IR-excess are Li-rich (A(Li) ≥ 1.0 dex), and importantly, three of them are SLRs with A(Li) ≥ 3.2 dex. The fraction of giants with IR excess among (23% of 13 SLRs in Sample 2) SLRs is significant. Among the Li-rich (A(Li) ≥ 1.0 dex) RC giants, the percentage of IR-excess giants is small, i.e. two out of 164 ( 1.2%), and none among Li-normal giants (A(Li) < 1.0 dex). The fraction of IR-excess stars among RC giants is similar to Martell et al. (2021b), who found two RC giants with A(Li) ≥ 2.7 dex which is significant (with a lower limit of about 15%) and comparable to ours in this study. They have not given how many of their reported 13 Li-rich giants (A(Li) > 1.5 dex as per their criterion) are SLRs in their RC sample. Recent study by Traven et al. (2020) for searching spectroscopic binaries among GALAH DR3 spectra showed that about 25% of giants having at least one evolved binary component. This is surprisingly in very good agreement with our IR excess SLRs fraction among total RC SLRs with mid-IR excess. Does this mean Li excess among RC giants is due to multiple mechanisms? One is due to only in-situ, for SLRs with no IR excess, and another, for SLRs with IR excess, is due to external events such as the direct addition of Li-rich material to the RC giants due to mergers (Casey et al. 2019; Aguilera-Gómez et al. 2016) or due to white dwarf triggered He-flash (Zhang & Jeffery 2013; Jeffery & Zhang 2020).

|  | RC | RGB | Total |
|---|---|---|---|
| With A(Li) | 957 | 518 | 1475 |
| with 2MASS (sample 1) | 957 | 518 | 1475 |
| With WISE (sample 2) | 418 | 359 | 777 |
| Sample 1 | | | |
| $v_{sini}$ | 957 | 518 | 1475 |
| rapid rotator | 168 | 73 | 241 |
| excess | 0 | 0 | 0 |
| Sample 2 | | | |
| $v_{sini}$ | 417 | 357 | 774 |
| rapid rotator | 53 | 64 | 117 |
| excess | 5 | 1 | 6 |

**Table 5.** $v_{sini}$ and excess statistics of the sample giants. Giants with $v_{sini} \geq 8$ km s$^{-1}$ are classified as rapid rotators.

**RC giants with no IR excess:** Most of the RC giants and the majority of SLRs in our sample do not have IR excess. At the RGB tip, Li enhancement in these giants may be due to the in-situ He-flash event. Theoretical models (Schwarzschild & Härm 1962; Thomas 1967) do not predict any observable effects on the outer surface, except a sudden drop in luminosity and corresponding contraction in size. Recent studies based on observational data showed substantial evidence that the He-flash at the RGB core holds the key for Li enhancement, and is common among RC giants (Kumar et al. 2020; Singh et al. 2021). Thus, the absence of IR excess in the majority of SLRs among RC giants is in accordance with the in-situ He-flash models .

**RC giants with IR excess:**

The models constructed by (Zhang & Jeffery 2013; Jeffery & Zhang 2020) predict Li enhancement due to the merger of white dwarfs with RGB giants triggering He-flash. Depending on the mass and the trajectory of a merging white dwarf with host giants on RGB, the core mass increases, leading to He-flash with subsequent Li abundance enhancement. Such mergers could happen anywhere on the RGB providing an additional channel for RC giant population. Assuming He-flash is the universal process for Li enhancement among RC giants, irrespective of the cause of He-flash event (in-situ or merger), and all the RC giants go through the SLR phase, one would expect a higher fraction of SLRs among RC giants selected based on measured IR colours compared to samples that were not selected based on IR colours. The fraction of SLRs among RC giants with mid-IR colours in our sample is about 3% which is a factor of 1.3 more than the SLR fraction among the total RC sample(22/957 or 2.3%, see table 4)). Further, the fraction of SLRs in our study is much higher, at least by a factor of 6 to 10, than the fraction of 0.11% SLRs among 17909 RC giants in the broader survey of Deepak & Reddy (2019) or 0.5% SLRs out of 4637 RCs in Singh et al. (2019) and 0.3% of SLRs out of 9284 RCs in Kumar et al. (2020). The higher fraction of SLRs in our study probably due to sample selection which has relatively smaller range in $T_{\rm eff}$ , and the sample is biased towards cooler temperatures (4500 K - 4800 K) compared to those much larger samples covering larger range in $T_{\rm eff}$ (4600 - 5200 K). What this means is that we have more younger RCs in our sample and hence more SLRs in which Li enhancement occurred very recently. Further, within our sample, fraction of SLRs in sample 2 RCs (with mid-IR colours) is more implying the SLRs in sample 2 RC are probably due to two channels: the SLRs with IR excess due to white dwarfs mergers with RGB cores and SLRs with no IR excess resulting from single star's evolution. This has another implication that the RC population may be a mixture of two kinds of giants; a majority one resulting from normal He-flash (without external events) in giants and the other a minority RC population resulting from mergers.

**IR excess and Li evolution among RC giants:**

It is known that the enhanced Li abundance post He-flash depletes rapidly with time as demonstrated by Singh et al. (2021). They have shown that the relation between asteroseismic properties ($\Delta P$) and A(Li) traces the time evolution of a central degenerate inert He-core to a fully convective He-burning core. Most of their SLRs are very young (with relatively low values of $\Delta P$) , and the Li-normal giants are relatively older RCs (with relatively high values of $\Delta P$). In between the two extremes are the Li-rich giants. Based on the predicted evolution of asteroseismic parameters and Li observations they suggested SLR phase may last for about 2 million years, beginning from the start of He ignition at the core. However, they could not determine whether the Li enhancement occurred during the main or immediate sub-flashes (within few thousand years post the RGB tip) due to a lack of reliable data in the low frequency (< 3-4 $\mu$Hz domain).

Our results of IR excess among giants corroborates rapid depletion of Li among RC giants. The results of no IR excess among Li-normal giants imply that most of the dust envelope evaporated by the time Li depleted to Li-rich or Li-normal giants. The number statistics provide estimation of SLR phase. Assuming RC phase lasts for about 100 Myrs (Deheuvels & Belkacem 2018) and the estimated SLR fraction of 0.11% from a large and well covered RC sample in $T_{\rm eff}$ by (Deepak & Reddy 2019) and 0.5% from the Kepler and LAMOST survey by (Singh et al. 2019) we calculate SLR phase ranging from 100,000 yrs to 500,000 yrs. However, the SLR fraction of 22% (22 SLRS out of 957 RCs), in this study, suggests SLR phase of about 2.2 Myrs. Since the sample in our current study is biased towards younger RCs (hence more SLRs), probably our value is overestimated. The duration of the He-flash and the subsequent Li-enhancement is in agreement with the upper limit of 2 Myrs set by (Kumar et al. 2020).

Further constraints on when the merger event, and hence the He-flash event took place come from the evolution of ejected mass. The estimated kinematic ages (*t*), the time it took for the dust envelope to reach a distance of inner radius of the dust envelop, given in table (see table 3) suggest merger event occurred very recently, a few hundred to a few thousand years ago. If this is true, Li-enhancement probably occurred during the main He-flash or immediately afterwards during first couple of sub-flashes. According to models the main flash seems to last for only a few thousand years (see Deheuvels & Belkacem (2018)). The theoretical model by Schwab (2020) does predict Li enhancement or SLRs during the main He-flash with subsequent rapid depletion of Li.

## CONCLUSION

We have searched for possible correlations between IR excess, Li enhancement and rotation among a well-defined sample of giants having well-measured Li abundances from GALAH, and IR fluxes from 2*MASS* and *WISE* surveys. The sample covers cool RC and upper RGB phases. Results show IR excess among giants is quite rare, just about 0.4% (six out of 1475 giants). However, among RC giants, the fraction is higher by a factor of three, i.e. about 1.2% (five out of 418 giants). Importantly, of the five RC giants with IR excess; three are super Li-rich, two are Li-rich, and none among 241 Li-normal giants suggesting giants with IR excess are more likely to be super Li-rich. Results suggest some kind of merger event, possibly with He-white dwarf, took place recently. Lack of IR excess among relatively older RC giants with normal Li suggests that by the time RCs evolved from SLR phase to Li-normal the ejected dust envelope evaporated. Further, the estimated kinematic ages ranging from a few hundred to a few thousand years for the IR excess giants agree with the estimated 100,000 yrs to 500,000 M yrs, based on number statistic, for the super Li-rich phase among RCs. This implies the cause for IR excess, a merger event, and the Li enhancement, the He-flash, occurred very recently. Dust modelling of giants with IR excess among RCs reveals asymmetric dust envelope morphology (opacity index $\beta$ either too high or too low), a signature of merger events with mass ejections. It would be a worthwhile exercise to analyse IR excess giants with polarimetric data to get further evidence of asymmetric distribution of dust envelopes, possibly caused by the mergers. Our analysis also reveals no particular trend of *vsini* values with the giants. However, we found the probability of finding rapid rotators among super Li-rich giants is twice higher than the Li-normal or Li-rich giants, and it is five to six times higher among Li-rich giants with IR excess.

## 5 ACKNOWLEDGMENTS

Authors thank Raghubar Singh for his helpful suggestions. This work has used data from GALAH DR3 survey acquired through https://www.galah-survey.org/dr3/the_catalogues/ and European Space Agency (ESA) mission Gaia (https://www.cosmos.esa.int/gaia). This publication also makes use of VOSA, developed under the Spanish Virtual Observatory project supported by the Spanish MINECO through grant AyA2017-84089, SIMBAD database, NASA ADS service, Castelli and Kurucz 2004 Stellar Atmosphere Models and DUSTY code developed by Ivezic et al. Authors thank the referee for making a number of suggestions which improved the manuscript.

## DATA AVAILABILITY

The data underlying this article were accessed from [GALAH DR3, https://t.co/UMgXFMNTqv?amp=1]. The derived data generated in this research will be shared on reasonable request to the corresponding author.

## REFERENCES

Aguilera-Gómez C., Chanamé J., Pinsonneault M. H., Carlberg J. K., 2016, ApJ, 829, 127
Alonso A., Arribas S., Martinez-Roger C., 1994, A&AS, 107, 365
Andrae R., et al., 2018, A&A, 616, A8
Bedding T. R., et al., 2011, Nature, 471, 608
Bharat Kumar Y., Reddy B. E., Muthumariappan C., Zhao G., 2015, A&A, 577, A10
Buder S., et al., 2021, MNRAS, 506, 150
Carlberg J. K., Smith V. V., Cunha K., Carpenter K. G., 2016, ApJ, 818, 25
Carney B. W., Latham D. W., Stefanik R. P., Laird J. B., 2008, AJ, 135, 196
Casey A. R., et al., 2019, ApJ, 880, 125
Castellani M., Castellani V., 1993, ApJ, 407, 649
Ceillier T., et al., 2017, A&A, 605, A111
Chan S. J., Kwok S., 1988, ApJ, 334, 362
Charbonnel C., Balachandran S. C., 2000, A&A, 359, 563
Charbonnel C., Lagarde N., 2010, A&A, 522, A10
Cutri R. M., et al., 2003, VizieR Online Data Catalog, p. II/246
Cutri R. M., et al., 2021, VizieR Online Data Catalog, p. II/328
Deepak Reddy B. E., 2019, MNRAS, 484, 2000
Deheuvels S., Belkacem K., 2018, A&A, 620, A43
Draine B. T., 2006, ApJ, 636, 1114
Drake N. A., de la Reza R., da Silva L., Lambert D. L., 2002, AJ, 123, 2703
Gaia Collaboration et al., 2018, A&A, 616, A1
Iben Icko J., 1967, ApJ, 147, 624
Ivezic Z., Nenkova M., Elitzur M., 1999, arXiv e-prints, pp astro–ph/9910475
Jasniewicz G., Parthasarathy M., de Laverny P., Thévenin F., 1999, A&A, 342, 831
Jeffery S., Zhang X., 2020, arXiv e-prints, p. arXiv:2011.02209
Kamath D., Wood P. R., Van Winckel H., Nie J. D., 2016, A&A, 586, L5
Kumar Y. B., Reddy B. E., Campbell S. W., Maben S., Zhao G., Ting Y.-S., 2020, Nature Astronomy, 4, 1059
Lind K., Asplund M., Barklem P. S., 2009, A&A, 503, 541
Lucey M., Ting Y.-S., Ramachandra N. S., Hawkins K., 2020, MNRAS, 495, 3087
Lynden-Bell D., Pringle J. E., 1974, MNRAS, 168, 603
Magrini L., et al., 2021a, A&A, 651, A84
Magrini L., et al., 2021b, A&A, 655, A23
Martell S. L., et al., 2021a, Monthly Notices of the Royal Astronomical Society
Martell S. L., et al., 2021b, MNRAS, 505, 5340
Mori K., Kusakabe M., Balantekin A. B., Kajino T., Famiano M. A., 2021, MNRAS, 503, 2746
Muthumariappan C., Kwok S., Volk K., 2006, ApJ, 640, 353
Rebull L. M., et al., 2015, AJ, 150, 123
Reddy B. E., Lambert D. L., 2005, AJ, 129, 2831
Romano D., et al., 2021, A&A, 653, A72
Schröder K. P., Cuntz M., 2005, ApJ, 630, L73
Schwab J., 2020, ApJ, 901, L18
Schwarzschild M., Härm R., 1962, ApJ, 136, 158
Singh R., Reddy B. E., Bharat Kumar Y., Antia H. M., 2019, ApJ, 878, L21
Singh R., Reddy B. E., Campbell S. W., Kumar Y. B., Vrard M., 2021, ApJ, 913, L4
Takeda Y., Tajitsu A., 2017, PASJ, 69, 74

Thomas H. C., 1967, Z. Astrophys., 67, 420  
Ting Y.-S., Hawkins K., Rix H.-W., 2018, ApJ, 858, L7  
Traven G., et al., 2020, A&A, 638, A145  
Vrard M., Mosser B., Samadi R., 2016, A&A, 588, A87  
Wallerstein G., Sneden C., 1982, ApJ, 255, 577  
Wright N. J., et al., 2008, MNRAS, 390, 929  
Zhang X., Jeffery C. S., 2013, MNRAS, 430, 2113  
de La Reza R., Drake N. A., da Silva L., 1996, ApJ, 456, L115  

This paper has been typeset from a TeX/LaTeX file prepared by the author.